\documentclass[11pt]{article}

\usepackage{mathptmx}

\usepackage{amsmath,amssymb} 

\newtheorem{remark}{Remark}[section]

\newtheorem{lemma}{Lemma}[section]

\newtheorem{theorem}{Theorem}[section]
\newtheorem{proposition}{Proposition}[section]

\newcommand{\proof}{\textsc{proof}\quad}
\newcommand{\qed}{\hfill \textsc{qed}}

\numberwithin{equation}{section}

\begin{document}
\title{Mathematical quantum Yang-Mills theory revisited}
 
\author{Alexander  Dynin\\
\textit{\small   Professor  Emeritus of Mathematics, Ohio State University,}\\
\textit{\small  Columbus, OH 43210, USA}, \textit{\small  dynin@math.ohio-state.edu}}

\maketitle 

\begin{abstract}
A   mathematically rigorous  relativistic    quantum Yang-Mills theory with an arbitrary semisimple  compact gauge  Lie group  is set up  in the Hamiltonian  canonical  formalism.  The theory is non-perturbative, without cut-offs, and agrees with the causality and stability principles. This paper  presents a fully revised, simplified, and corrected version  of the corresponding material in  the previous papers \textsc{\small dynin}\cite{Dynin-1} and \cite{Dynin-2}.   The principal result  is established anew:  due to the quartic  self-interaction term in the Yang-Mills Lagrangian along with  the semisimplicity of the gauge group, the  quantum Yang-Mills  energy spectrum has a  positive   mass gap.  Furthermore,  the quantum Yang-Mills Hamiltonian has a countable orthogonal eigenbasis in a Fock space, so that the quantum Yang-Mills  spectrum is point and countable. In addition a fine structure of the spectrum is elucidated.  

\smallskip
\textsc{keys}: Millennium Yang-Mills problem; Finite propagation speed; Sobolev inequalities; Nuclear vector spaces; Infinite-dimensional holomorphy;   Friedrichs operator  extensions; Variational spectral principle; Symbols and spectral theory of pseudo-differential operators.

\medskip
2010  AMS Subject Clasification: 81T13, 81T07.
\end{abstract}

\section{Introduction}
\subsection{Context}
I  address \emph{both items}  of the Clay Mathematics Institute "Quantum Yang-Mills theory" problem requiring  a mathematical proof  that
\begin{quotation}
 for any compact semisimple global gauge group, a nontrivial  quantum Yang-Mills theory exists on the four-dimensional Minkowski spacetime and has a positive mass gap.    Existence includes establishing
 axiomatic properties at least as strong as G\r{a}rding-Whigtman axioms (see \textsc{jaffe-witten}\cite{A}).
\end{quotation}
\emph{As such this is a  problem of mathematical existence}.  It   does not require  a reconstruction of the conventional quantum Yang-Mills theory. (Notably the  famous  LHC experimental discovery of a "Higgs scalar  field" has not verified  the hypothetical Higgs mechanism for the origin of a positive mass  of classical Yang-Mills fields.)

The proposed  mathematically rigorous  quantum Yang-Mills theory  is relativistic, non-perturbative, and constructive. It does imply a positive mass gap in the spectrum of quantum Yang-Mills Hamiltonian.
Actually the whole spectrum is described  qualitatively.

\medskip

 Mathematical foundations of  quantum mechanics are J. von Neumann  theory of (unbounded) Hermitian  operators in a Hilbert space and  H. Weyl canonical quantization rule. The latter has  led to E. Wigner quantum statistical mechanics
and then to calculus of annihilation and creation operators in quantum optics (see \textsc{\small agarval-wolf}\cite{Agarwal}) on the physics side, and to theory of pseudo-diffefential operators (see \textsc{\small shubin}\cite[Chapter 4]{Shubin} and \textsc{\small folland}\cite[Chapter 2]{Folland}) on the mathematics side.

The symbolic calculus is a far reaching generalization of the classical Heaviside  operational calculus for ordinary differential equations. 
The \emph{non-linear} Hamiltonian function on the phase space of a classical mechanical system in an Euclidian space may serve e.g. as normal, or  Weyl,  or anti-normal symbol of the \emph{linear} Schr\"{o}dinger partial differential operator for a quantum analogue of the system. The   symbol choice depends on the ordering of  annihilation and creation operators or equivalently of operators of multiplication and partial differentiation in  the Schr\"{o}dinger operator.

\medskip

In quantum field theory the first infinite-dimensional functional   Schr\"{o}dinger equation  was introduced by  P. Jordan and W. Pauli (\emph{Zur Quantumelectrodynamik ladungsfreier Felder,}  Zeitung  f\"{u}r Physik, \textbf{47} (1928)). Much later such operators have been used  by J. Schwinger during 1950's. Yet
\begin{quotation}
"Mathematically, quantum field theory involves integration, and 
elliptic operators, on infinite-dimensional spaces. Naive attempts 
to formulate such notions in infinite dimensions lead to all sorts of 
trouble. To get somewhere, one needs the very delicate constructions
considered in physics, constructions that at first sight look rather 
specialized to many mathematicians. For this reason, together with 
inherent analytical difficulties that the subject presents,
rigorous understanding has tended to lag behind development of physics."
\textsc{\small witten}\cite[p.346]{Witten}
\end{quotation} 
In 1954  \textsc{gelfand-minlos}\cite{Gelfand} proposed to solve Schwinger infinite-dimensional  partial differential  equations  via approximations by  finite-dimensional ones with large number of independent variables (see \textsc{berezin}\cite[Preface]{Berezin-65}). Such  approximations drastically differ from the customary lattice approximations. Afterwards   a  rigorous mathematics of infinite-dimensional  partial differential operators has been developed   along Gelfand-Minlos lines,  in particular, by   \textsc{\small kree-raczka}\cite{Kree78} in a cylindrical formalism.  Simultaneously P. Kree  found an  alternative formalism  of   Gelfand nuclear triples  (see \textsc{\small kree}\cite{Kree77}). It enables  a mathematically rigorous symbolic calculus  similar to the heuristic real-analytic Agarwal-Wolf calculus (see \textsc{\small agarwal-wolf}\cite{Agarwal})   in quantum optics.   An important difference with the standard symbolic calculus of finite-dimensional  pseudo-differential operators   is that    it is based on convergent series expansions   rather than  asymptotic ones. 
In general terms, Martineau analytic functionals replace Schwartz distributions.  This is a natural frame-work
for infinite-dimensional generalization of Agarwal-Wolf calculus of creation and annihilation operators.

\subsection{Issues}
\subsubsection{Conventions}
We use natural units in quantum field theory:  Planck's  $\hbar$ (relevant for quantum effects), Einstein's $c$ (relevant  for relativistic effects), and the energy  unit GeV, or the reciprocal  length fermi unit fm (relevant for nuclear physics). 

Dimensional homogeneity is maintained carefully. In particular Fock spaces are built over Hilbert
spaces with dimensionless scalar products.

The scaleless Yang-Mills coupling  constant is suppressed. 

\subsubsection{Classical Yang-Mills fields} 
  In  the global Hamiltonian (aka temporal) gauge relativistic Yang-Mills equations are known to form a non-linear    hyperbolic system of the 2nd order partial  differential equations on the Minkowski space    $\mathbf{R}^{1,3}$ with the finite propagation speed $\leq 1$ property of solutions dubbed classical Yang-Mills fields.  The relativistic invariance of the Yang-Mills Lagrangian implies the relativistic covariance of Yang-Mills fields.
  
   In the 1st order formalism the Yang -Mills equations form an infinite-dimensional Hamiltonian system (see \textsc{\small faddeev-slavnov}\cite[Section III.2]{Faddeev}). The  Hamiltonian equations  have a unique global solution  for the Cauchy problem with given  initial data constrained by dynamically invariant non-linear partial differential equations  (see \textsc{\small goganov-kapitanskii}\cite{Goganov}). Because of  the finite propagation speed,  global solutions are  generated  by the solutions with  the   constrained  initial data restricted to the central  balls $\mathbb{B}=\mathbb{B}(r)$ of the radius $0<r<\infty$ in $\mathbb{R}^3$.  
  
In general, initial  Cauchy data may be assigned on any space-like hyperplane in Minkowski space with Lorentz orthogonal time axis presumed to be future oriented. Since the proper Lorentz transformations act transitively on the  future oriented time-like causal cone, the assigned Cauchy problems are relativistically equivalent.

\subsubsection{Classical  Yang-Mills Hamiltonian}
The relativistic invariance of the  classical Yang-Mills action functional of  solutions with  compactly supported intial data implies  Noether energy-momentum 4-vector $P=(P^\mu)$ on  Minkowski space. The functional  time-component $P^0$ is the  classical Yang-Mills Hamiltonian and the   Euclidean  3-vector $(P^1,P^2,P^3))$ is the Yang-Mills  momentum functional.

Constrained initial data with compact supports  generate all solutions of Yang-Mills equations and  the classical    Yang-Mills Hamiltonian  is completely defined by them.   Hence   it is  uniquely defined by classical Yang-Mills fields  compactly  supported by  balls   $\mathbb{B}$.

 By the coordinate scaling invariance of the Yang-Mills action functional, the energy functional   $P^0$ is inversely proportional to $r$.   The  functional $H:=rP^0$ is the  \emph{scaleless  Yang-Mills   Hamiltonian}. Since $H$ is invariant under scaling transformation, it is the same on all balls $\mathbb{B}(r)$.

The time-independent   gauge invariance of Yang-Mills   Hamiltonian $H$ in the temporal gauge allows to reduce the non-linear phase space of constrained initial data to  infinite-dimensional  \emph{vector space  of   transversal initial data} in   
$\mathbb{B}$.   Thus    the calculus of operators of creation and annihilation in a Fock space  over a Hilbert space of  such transversal data is used for a  quantum Yang-Mills theory.

\subsubsection{Canonical quantization}

\begin{quotation}
Classical mechanics has provided such a successful framework in physics that it is natural to rephrase physical systems in terms of  fixed time degree of freedom which evolve in time. 
(see \textsc{\small reed-simon}\cite[Page 215]{Reed2}.
\end{quotation} 
 Hamiltonian formulation of classical Yang-Mills theory (see  \textsc{\small faddeev-slavnov}\cite[Chapter III, Equations (2.64)]{Faddeev}) is such a paraphrase. Its relativistic covariance  is implied by the relativistic covariance of the equivalent Yang-Mills equations.

The Hilbert spaces $\mathcal{L}^2_\bot(\mathbb{B}(r))$  of transversal initial data with the scaleless inner products are scaling invariant, and so are the rigged Fock spaces over them. 

The scaleless Yang-Mills Hamiltonian functional is considered as  the \emph{anti-normal symbol} of the
 scaleless quantum Yang-Mills Hamiltonians in a rigged Fock spaces  over transversal initial data in  
 $\mathbb{B}(r)$. The operators are densely defined  in the Fock space over 
$\mathcal{L}^2_\bot(\mathbb{B}(r))$ and  are unitarily equivalent.

Since scaleless Yang-Mills Hamiltonian is non-negative, the proposed  quantum Yang-Mills Hamiltonians have  unique non-negative Friedrichs   operator extensions \footnote{ By \textsc{\small glimm-jaffe}\cite{Glimm69}, the normal quantization of a non-negative functional is not necessary  a non-negative operator.}  By the unitary equivalence the operators have the same spectrum, the Yang-Mills spectrum of the paper title. 

Due to gauge invariance  Yang-Mills Hamiltonian does not contain a positive quadratic form term. However,    the \emph{quartic term} of the Hamiltonian functional along with the semisimplicity of the gauge group entail such quadratic form in the  Weyl symbol of  quantum Yang-Mills Hamiltonian. \footnote{ This   argument  cannot be applied  to the  photonic Maxwell-Schr\"{o}dinger operator!} 
The arising mass  quadratic form implies that the Friedrichs Hermitian  extension of  quantum Yang-Mills Hamiltonian   has a countable orthonormall eigenbasis in a Fock space, so that the quantum Yang-Mills  spectrum is point and countable. 

 It is shown that monomial  multiparticle eigen-states form a countable  eigenbasis in the Fock Hilbert space, so that the Yang-Mills quantum energy-mass spectrum is a countable set of eigenvalues. Furthermore, there is a positive mass gap at the spectrum bottom.

Since  scaleless quantum Yang-Mills Hamiltonian is not relativistically invariant, the Yang-Mills mass depends on Lorentz coordinate frame (as in classical special relativity).
Restoration the physical dimension $[L]^{-1}$ of    Yang-Mills Hamiltonian implies that the mass gap is proportional to the  classical energy level.   

\subsection{Acknowledgements.}

 I am thankful  to  Clifford Taubes for warning that  Yang-Mills Hamiltonian was oversimplified
in \textsc{\small dynin}\cite{Dynin-1} and \textsc{\small dynin}\cite{Dynin-2}. The error is corrected in the present paper.

I am grateful to  L. D. Faddeev for useful discussions  and to  M. Frasca for moral support. 
 
\section{Analysis in Bargmann-Fock space}
Basic  references are \textsc{\small kree}\cite{Kree}, \textsc{\small kree}\cite{Kree77}, and 
\textsc{\small kree-raczka}\cite{Kree78}.

\subsection{P. Kree  rigging of a Bargmann-Fock space}
The compexification $\mathcal{H}^0:=\mathcal{X}\oplus i\mathcal{X}$ of a separable real Hilbert space 
$\mathcal{X}$ carries the complex conjugation $z:=x+iy\mapsto  \overline{z}:=x-iy$, an anti-linear isometric involution. 

 The conjugation converts the Hilbert  space  $\mathcal{H}^0$   into the  anti-dual space 
 $\overline{\mathcal{H}^0}$.  The   Hermitian  scalar  product of $\overline{z}$ and $w$ in   
$\mathcal{H}^0$ is denoted by $\overline{z}w$ (as in $\mathbb{C}$), a shorthand for the Dirac's $\langle z|w\rangle$.  

\smallskip
A  \emph{nuclear Gelfand sesqui-linear rigging} of the  Hilbert space $\mathcal{H}^0$ is  a  triple  of dense    continuous embeddings (see  \textsc{\small  gelfand-vilenkin}\cite{Gelfand-1})\\
\begin{equation}
\label{eq:Gelfand}
\mathcal{H}\ \subset\ \mathcal{H}^0\ \simeq\  \overline{\mathcal{H}^0}\  \subset\ \mathcal{H}^*, 
\end{equation} 
where 
\begin{itemize}
\item   A nuclear countably Hilbert  space $\mathcal{H}$  is   the  intersection of a countable nested family of Hilbert spaces
\begin{equation*}
\bigcap \mathcal{H}^n,\ n\geq 0,\ \mathcal{H}^{n+1}\subset\mathcal{H}^n,
\end{equation*}
  where the embedding are nuclear linear operators  with  dense  ranges  (see   \textsc{\small gelfand-vilenkin}\cite{Gelfand}). The topology is defined by the simultaneous convergence in all $\mathcal{H}^n$. In fact 
  $\mathcal{H}$ is a Frechet nuclear space (see \textsc{\small treves}\cite{Treves}).
 
  \item The strong anti-dual  $\mathcal{H}^*$ of $\mathcal{H}$ of continuous  anti-linear functionals $z^*w:=z^*(w)$ on  $\mathcal{H}^n$  of continuous  anti-linear functionals $z^*w:=z^*(w)$ is  the  union  of the anti-dual Hilbert spaces $\mathcal{H}^{n*}$ of $\mathcal{H}^n$ with the topology defined by convergence in a $\mathcal{H}^{n*}$.  In fact $\mathcal{H}^*$ is a nuclear  LF-space (see \textsc{\small treves}\cite{Treves}).
  
  \item The equivalence $\mathcal{H}^0\ \simeq\  \overline{\mathcal{H}^0}$ is defined  via Riesz representation of anti-linear functionals (see \textsc{\small treves}\cite{Treves}). In particular, $z^*=\overline{z}$ for $z\in\mathcal{H}^0$.
 \end{itemize}

The Bargmann-Fock space $\mathcal{K}(\mathcal{H}^0)$ is  the   Hilbert space  of  entire analytic functionals
$\Psi(\zeta
^*)$ on $\mathcal{H}^*$ that are  square integrable 
with respect to  the  Gaussian probability  measure   
  $d\gamma$  on  $\mathcal{H}^*$. The latter is   uniquely defined through its  pull-offs via finite rank Hermitian   projectors $p:\mathcal{H}^*\rightarrow\mathcal{H}$ from the finite-dimensional Gaussian probability measures  on finite-dimensional complex subspaces $p(\mathcal{H}^*)$ 
  \begin{eqnarray}
   & &
   \label{eq:gaussian}
d\gamma(p\overline{\zeta
},p \zeta
)\ =\    c(p)^{-1}d(p\overline{\zeta
})(dp  \zeta
)e^{-\overline{p\zeta
}p\zeta
},\\ 
   & &
\nonumber c(p):=\int_{p\mathcal{H}^*}d\gamma(p\overline{\zeta
},p\zeta
)e^{-\overline{p\zeta
}p\zeta
}
   \end{eqnarray}  
 (see \textsc{\small  gelfand-vilenkin}\cite{Gelfand-1}).  

The complex conjugation  in $\mathcal{K}(\mathcal{H}^0)$ is   $\overline{\Psi}(\zeta):=\overline{\Psi^(\overline{\zeta})}$, and the Hermitian scalar product is
\begin{equation*}
\langle\ \overline{\Psi} |\ \Phi\ \rangle\ :=\ \int_{\mathcal{H}^*}\!d\gamma\:\overline{\Psi}(\zeta
)\Phi(\zeta
^*).
\end{equation*}
In view of  the Fernique theorem (see \textsc{\small kuo}\cite[Chapter 3, Theorem 2.4]{Kuo}, the  functionals are integrable with respect to  the  Gaussian  measure on $\mathcal{H}^*$. \footnote{ The Fernique theorem per se involves the Wiener space $\mathcal{H}^{1*}$, a carrier of Gaussian probability measure    $d\gamma$ in  the bigger  space $\mathcal{H}^*$.} 

The \emph{Kree rigging} of the  Fock space $\mathcal{K}(\mathcal{H}^0)$ consists of the dense  continuous embeddings of complex topological vector  spaces (see  \textsc{\small  kree}\cite{Kree77} and \cite[Subsection (5.9]{Kree78})
\begin{equation}
\label{eq:Kree}
 \mathcal{K}(\mathcal{H}^*)\ \subset\  \mathcal{K}(\mathcal{H}^0)\  \subset\ \mathcal{K}(\mathcal{H}),
\end{equation}
where 
\begin{itemize}
   \item     The nuclear countably-Hilbert  space  $\mathcal{K}(\mathcal{H}^*)$ of the  space of   entire holomorphic  functionals   
$\Phi(z^*)$ on $\mathcal{H}^*$ of the first order exponential growth on every 
$\mathcal{H}^{n*}$  \footnote{ A functional on $\mathcal{H}$ is \emph{holomorphic} if it is holomorphic on every  finite-dimensonal subspace  and is continuous on every $\mathcal{H}^{n*}$.  Since   $\mathcal{H}^*$ is a nuclear Silva space, this property is equivalent to  the Silva-analyticity  in \textsc{\small colombeau}\cite[Chapter 2]{Colombeau}.}.   
     
    By  \textsc{\small boland}\cite{Boland} and \textsc{\small colombeau}\cite[Chapter 8. Abstract]{Colombeau} the countably Hilbert  space $\mathcal{K}(\mathcal{H})$ is nuclear. In particular, it is reflexive.
    
     \item    The space $\mathcal{K}(\mathcal{H})$ is the space of all continuous Gateaux  holomorphic functionals $\Psi(w)$ on  $\mathcal{H}$   with the topology of uniform convergence on compact subsets of $\mathcal{H}$. 

By    \textsc{\small boland}\cite{Boland}, the space
 $\mathcal{K}(\mathcal{H})$ is the strong dual of  $\mathcal{K}(\mathcal{H}^*)$.  Furthermore $\mathcal{K}(\mathcal{H})$ is a nuclear space (see  \textsc{\small treves}\cite[Proposition 50.6]{Treves}).
\end{itemize}

\subsection{Borel transform}       
   \emph{Coherent (aka exponential) states} $e^z(\zeta^*):=e^{\zeta^*z},\ z\in\mathcal{H},$ have  the following well known basic properties \footnote{ They are straightforward on cylindrical states $p\zeta^*$ and then, by strong limits, are extended to all states. Note also that  the coherent states are cylindrical.}
\begin{itemize}
\item Any $\Psi\in\mathcal{K}(\mathcal{H}^*)$ where it has a unique coherent states expansion 
\begin{equation}                                                                                            
\label{eq:reproducing}
\Phi(\overline{\zeta})\ =\ \int_{\mathcal{K}(\mathcal{H}^*)}d\gamma(\zeta^*,\zeta)\: \overline{e^{z}(\zeta^*)}\Phi(\zeta^*).
\end{equation}
This equation is for $\Phi(z^*)$ on the dense  the subspace $\mathcal{H}\subset\mathcal{H}^*$ where 
$z^*=\overline{z}$ that defines $\Phi(z^*)$ uniquely.

\item   \emph{Overlap identity}
\begin{equation}                                                                                            
\label{eq:overlap}
\langle\ e^{z}\ |\ e^{w}\ \rangle\ =\ e^{\overline{z}w}, \quad z,w\in\mathcal{H}.
\end{equation}
\proof
It follows from  \textsc{\small folland}\cite[Chapter 1, theorem (1.63)]{Folland} that  for non-negative integers 
$m,n$ we have $\langle z^{n}|w^n\rangle=n!(\overline{z}w)^n$ and that $z^m|w^n\rangle=0$ if  $m\neq n$.
Then
\begin{equation*}
\langle e^{z}\ |\ e^{w}\ \rangle\ =\ \langle\ \sum_{m=0}^\infty z^{m}/m!\ | \ \sum_{n=0}^\infty w^n/n!\ \rangle\ =\ e^{\overline{z}w}. 
\end{equation*}
\qed

\item    \emph{Borel transform}  $F$  (see \textsc{\small  colombeau}\cite[Chapter 7]{Colombeau},  as well as  \textsc{\small treves}\cite[chapter 22]{Treves})  is
 \begin{equation}                                                                                            
 \label{eq:basis}
\Phi(\overline{\zeta})\mapsto (F\Phi)(\overline{z})\ :=\ \langle\ e^{z}\  | \ \Phi \ \rangle,\quad \Psi(\zeta)\mapsto (F\Psi)(z)\  :=\  \langle\ \Psi\  |\ e^{z}\ \rangle.
\end{equation}
By (\ref{eq:overlap}) the Borel transform $F(e^{w}(\overline{\zeta^*}))=e^w(\overline{z^*}))$. Thus
 Borel transform induces  a linear topological  isomorphism  of the corresponding riggings of the  Fock space (see  \textsc{\small kree}\cite[Theorem (2.15)]{Kree77}.
\item  Borel transform  intertwines  the directional differentiation 
$\partial_{\overline{z}},\ z\in\mathcal{H},$   and the multiplication with $\overline{\zeta}z$ in  $\mathcal{K}(\mathcal{H})$.

\item \emph{Sesqui-linear Borel transform} in the sesqui-linear Kree triple
\begin{equation}
\label{eq:sesqui}
\mathcal{K}(\overline{\mathcal{H}^*}\times\mathcal{H}^*)\ \subset\ 
\mathcal{K}(\overline{\mathcal{H}^0}\times\mathcal{H}^0)\ \subset\ 
\mathcal{K}\overline{\mathcal{H}}\times\mathcal{H})
\end{equation}
is tensor product of Borel transforms defined by sesqui-holomorphic coherent states
$e^{\overline{\eta_1^*z_1}+\eta_2^*z_2}\ =\ e^{\overline{z_1}}(\overline{\eta_1^*})e^{z_1}(\eta_1^*)$.
\end{itemize}

\subsection{Calculus of  creation and annihilation operators}
 \emph{Operators of creation and annihilation} are continuous operators of multiplication and complex directional differentiation in $\mathcal{K}(\mathcal{H}^*)$ and $\mathcal{K}(\mathcal{H})$,
\begin{eqnarray}
& &                                                                                           
\label{eq:creators}
\widehat{z}\:\Phi(\zeta^*)\  :=\  (\zeta^*z)\Phi(\zeta^*),\quad
\widehat{z^*}\Phi(\zeta^*)\ :=\ \partial_{z^*}\Phi(\zeta^*),\\
& &
\label{eq:cocreators} 
\widehat{z}\:\Psi(\zeta)\  :=\ \partial_{z}\Psi(\eta)),\quad \widehat{z^*}\Psi(\eta)\ :=\ 
(z^*\eta)\Psi(\eta)
\end{eqnarray}   
\begin{itemize}
\item The adjoint of a creation operator $\widehat{z}$ is the annihilation operator $\widehat{\overline{z}}$.
 \item Operators $\widehat{z}$ and  $\widehat{\overline{z}}$ are continuos in $\mathcal{K}(\mathcal{H})$ and
 $\mathcal{K}(\mathcal{H})^*$ where they generatev.
 \item  Canonical bosonic commutation  relations (CCR) take the form of  
  \begin{equation}                                                                                           
  \label{eq:CCR}
[\widehat{z_1^*},\widehat{z_2}]\ =\ (z_1^*z_2),\quad [\widehat{z_1^*},\widehat{z_2}]\
=\ 0, \quad [\widehat{z_1},\widehat{z_2}]\ =\ 0.
\end{equation}
\item  \emph{Coherent states}  $e^{z}$  are the eigenstates of the annihilation      
 operators: $\widehat{z^*}e^{w}\ =\ (z^*w)e^z$.
\item Creators and annihilators generate  strongly continuous commutative  operator groups in 
$\mathcal{K}(\mathcal{H}^*)$ and $\mathcal{K}(\mathcal{H})$
\begin{eqnarray}
& &
e^{\widehat{z}}\Phi(\zeta^*)\ =\ e^{w^*z}\ \Phi(\zeta^*),\quad e^{\widehat{z^*}}\Phi(\zeta^*)\ =\  \Phi(\zeta^*+z^*),\\
& &
e^{\widehat{z}}\Psi(\zeta)\ =\ \Psi(\zeta+z)\quad e^{\widehat{z^*}}\Psi(\zeta)\ =\  e^{\zeta^*z}\Psi(\zeta).
\end{eqnarray}
The  operator products 
$e^{\widehat{z}}e^{\widehat{z^*}}$  and $e^{\widehat{z*}}e^{\widehat{z}}$ 
are invertible continuous operators in $\mathcal{K}(\mathcal{H})$ and $\mathcal{K}(\mathcal{H}^*)$.
\end{itemize}

By Baker-Campbell-Hausdorff  formula and the canonical  commutation  relations (\ref{eq:CCR}),
 \begin{equation}                                                                                           
 \label{eq:BCH}
e^{\widehat{z}}e^{\widehat{z^*}}\ =\ e^{\widehat{z}+\widehat{z^*}}
e^{z^*z/2}, \quad  e^{\widehat{z^*}}e^{\widehat{z}}\ =\ 
e^{\widehat{z}+\widehat{z^*}}e^{-z^*z/2}.
\end{equation}
Therefore  the operator $e^{\widehat{z}+\widehat{z^*}}$ is also continuous and invertible in 
 $\mathcal{K}(\mathcal{H})$.

The  sesqui-holomorphic Borel transform of
$\tilde{\Theta} \in\mathcal{K}(\overline{\mathcal{H}}\times\mathcal{H})$
\begin{equation}
\label{eq:Theta}
	\Theta(\overline{z_1},z_2)\ =\  \langle \tilde{\Theta}(\overline{\eta_1},\eta_2)\ |\ e^{\eta_1}(\overline{z_1})e^{\overline{\eta_2}}(z_2)\ \rangle
\end{equation}
is quantized as the continuous \emph{normal, Weyl, and anti-normal operators} from 
$\mathcal{K}(\mathcal{H}^*)$ to $\mathcal{K}(\mathcal{H})$ defined by  the corresponding normal, Weyl, and anti-normal ordering of creators and annihilators
\begin{eqnarray}
& &
\label{eq:n} 
\widehat{\Theta}_\nu\ =\ \Theta_\nu(\widehat{\overline{\eta_1}}, \widehat{\eta_2})
\  :=\ 
 \big\langle\ \tilde{\Theta}_\nu(\overline{\eta_1},\eta_2)\ \big| e^{\widehat{\eta_1}}e^{\widehat{\overline{\eta_2}}} \big\rangle,\\ 
& &
\label{eq:w}
\widehat{\Theta}_\omega\ =\ \Theta_\omega(\widehat{\overline{\eta_1}}, \widehat{\eta_2})
\ :=\ \big\langle\ \tilde{\Theta}_\omega(\eta_1,\overline{\eta_2}) \big|\ e^{\widehat{\eta_1}+\widehat{\overline{\eta_2}}}\ \big\rangle,\\ 
& &
\label{eq:a}
\widehat{\Theta}_\alpha\ =\ \Theta_\alpha(\widehat{\overline{\eta_1}}, \widehat{\eta_2})
\ :=\ \big\langle\ \tilde{\Theta}_\alpha(\eta_1 ,\overline{\eta_2})\big |\ e^{\widehat{\overline{\eta_2}}}e^{\widehat{\eta_1}}\ \big\rangle.
\end{eqnarray}
The coherent matrix elements
\begin{eqnarray*}
& &
\big\langle\ e^{z_1}\ \big|\ \widehat{\Theta}_\nu\ \big| e^{z_2}\ \big\rangle\ =\
 \big\langle\ \tilde{\Theta}\nu(\eta_1,\overline{\eta_2})\ \big |\  \langle\ e^{z_1}\ |  \  e^{\widehat{\eta_1}}e^{\widehat{\overline{\eta_2}}}\ | e^{z_2}\ \rangle \big\rangle,
 \\ 
& &
\big\langle\ e^{z_1}\ \big|\ \widehat{\Theta}_\omega\ \big| e^{z_2}\ \big\rangle\ =\
 \big\langle\ \tilde{\Theta}_\omega(\eta_1,\overline{\eta_2})\ \big |\  \langle\ e^{z_1}\ |  \ 
 e^{\widehat{\eta_1}+\widehat{\overline{\eta_2}}}\ | e^{z_2}\ \rangle \big\rangle,
 \\ 
& &
\big\langle\ e^{z_1}\ \big|\ \widehat{\Theta}_\alpha\ \big| e^{z_2}\ \big\rangle\ =\
 \big\langle\ \tilde{\Theta}_\alpha(\eta_1,\overline{\eta_2})\ \big |\  \langle\ e^{z_1}\ |  \  e^{\widehat{\overline{\eta_2}}}e^{\widehat{\eta_1}}\ | e^{z_2}\ \rangle \big\rangle
\end{eqnarray*}
are well defined. For starters
\begin{equation*}
  \langle\ e^{z_1}\ |  \  e^{\widehat{\eta_1}}e^{\widehat{\overline{\eta_2}}}\ | e^{z_2}\ \rangle\ =\ 
 \langle\ e^{\widehat{\overline{\eta_1}}}e^{z_1}\  |\ e^{\widehat{\overline{\eta_2}}} e^{z_2}\ \rangle\ =\ \langle\ e^{\overline{\eta_1}z_1}e^{z_1}\ |\ e^{\overline{\eta_2}z_2}e^{z_2}\ \rangle\
\stackrel{(\ref{eq:overlap})}{=}\ e^{\overline{z_1}\eta_1+\overline{\eta_2}z_2}e^{\overline{z_1}z_2}
\end{equation*}
imply via (\ref{eq:Theta}) that
\begin{equation}
\label{eq:n1}
\big\langle\ e^{z_1}\ \big|\ \widehat{\Theta}_\nu\ \big| e^{z_2}\ \big\rangle\ =\   \langle \tilde{\Theta}_\nu
(\overline{\eta_1},\eta_2)\ |\ e^{\eta_1}(\overline{z_1})e^{\overline{\eta_2}}(z_2)\  e^{\overline{z_1}z_2}\ \rangle\ =\ \Theta_\nu(\overline{z_1},z_2)e^{\overline{z_1}z_2}.
\end{equation}
Since $\Theta_\nu(\overline{z_1},z_2)e^{\overline{z_1}z_2}$ is sesqui-holomorphic on  $\overline{\mathcal{H}}\times\mathcal{H}$, 
the normal coherent state matrix 
$\langle\ e^{z_1}\ |  \  e^{\widehat{\eta_1}}e^{\widehat{\overline{\eta_2}}}\ | e^{z_2}\rangle$ is the Grothendieck kernel
of a continuous linear operator $\widehat{\Theta}_\nu:\ \mathcal{K}(\mathcal{H}^*)\rightarrow \mathcal{K}(\mathcal{H})$ (see  \textsc{\small treves}\cite{Treves}).

Vice versa for any continuous linear operator $Q:\  \mathcal{K}(\mathcal{H}^*)\rightarrow \mathcal{K}(\mathcal{H})$
the sesqui-holomorphic functional
\begin{equation}
\label{eq:n2}
\Theta_\nu^Q(\overline{z_1},z_2)\ :=\ \langle\ e^{z_1}\ | \ Q \ |\  e^{z_2}\ \rangle e^{-\overline{z_1}z_2}
\end{equation}
belongs to $\mathcal{K}(\overline{\mathcal{H}}\times\mathcal{H})$ and
satisfies
\begin{equation}
\langle\ e^{z_1}\ | \ Q \ | \  e^{z_2}\ \rangle\ =\ \big\langle\ \tilde{\Theta}_\nu^Q(\overline{z_1},\eta_2)\ \big |\  \langle\ e^{z_1}\ | \  e^{\widehat{\eta_1}}e^{\widehat{\overline{\eta_2}}}\ |\ e^{z_2}\ \rangle \big\rangle
\end{equation}
As in (\ref{eq:BCH}) 
\begin{equation}                                                                                           
 \label{eq:BCH1}
e^{\widehat{\eta_1}}e^{\widehat{\overline{\eta_2}}}\ =\ e^{\widehat{\eta_1}+\widehat{\overline{\eta_2}}}
e^{-\overline{\eta_2}\eta_1/2}, \quad  e^{\widehat{\overline{\eta_2}}}e^{\widehat{\eta_1}}\ =\ 
e^{\widehat{\eta_1}+\widehat{\overline{\eta_2}}}e^{-\overline{\eta_2}\eta_1/2},
\end{equation}
so that
\begin{equation}
\tilde{\Theta}_\nu(\overline{\eta_1},\eta_2) \ =\ e^{\overline{\eta_1}\eta_2/2}
\tilde{\Theta}_\omega(\overline{\eta_1},\eta_2), \quad 
\tilde{\Theta}_\omega(\overline{\eta_1},\eta_2)\ =\ 
e^{\overline{\eta_1}\eta_2/2}\tilde{\Theta}_\alpha(\overline{\eta_1,\eta_2}).
\end{equation}
Applying the intertwining property of sesqui-holomorhic Borel transform one gets
\begin{equation}
\label{eq:omeganu}
\Theta_\nu(\overline{z_1},z_2)\ =\ e^{\overline{\partial_{z_1}}\partial_{z_2}/2}\Theta_\omega(\overline{z_1},z_2), \quad
\Theta_\omega(\overline{z_1},z_2)\ =\ e^{\overline{\partial_{z_1}}\partial_{z_2}/2}
\Theta_\alpha(\overline{z_1},z_2).\end{equation}
These equations  define $\Theta_\omega^Q(\overline{z_1},z_2)$ and  $\Theta_\alpha^Q(\overline{z_1},z_2)$
via  $\Theta_nu^Q(\overline{z_1},z_2)$ for all continuous linear operators $Q:\ \mathcal{H}\rightarrow\mathcal{H}^*$.

A sesqui-holomorphic functional $\Theta(\overline{z_1},z_2)$ is uniquely defined by its restriction 
$\sigma(\overline{z},z)$    to the real diagonal   of $\overline{\mathcal{H}}\times\mathcal{H}$. The restrictions
\begin{equation}
\label{eq:symbols}
\sigma_\nu^Q(\overline{z},z),\quad \sigma_\omega^Q(\overline{z},z),\quad \sigma_\alpha^Q(\overline{z},z)\end{equation}
are the \emph{normal, Weyl, anti-normal symbols} of an operator $Q$.  The symbols are arbitrary real analytic functionals on the real diagonal of $\overline{\mathcal{H}}\times\mathcal{H})$. Furthermore, under the corresponding ordering of creators and annihilators,
\begin{equation}
\label{eq:operators}
Q=\sigma_\nu^Q(\widehat{\overline{z}},\widehat{z})\ =\  \sigma_\omega^Q(\widehat{\overline{z}},\widehat{z})\ =\ \sigma_\alpha^Q(\widehat{\overline{z}},\widehat{z}).
\end{equation}

 Formulas (\ref{eq:n2}) and (\ref{eq:operators}) reproduce well common formulas for finite-dimensional pseudo-differential operators (see \textsc{small folland}\cite[Chapter 3, Section 7]{Folland}).

In terms of their  matrix elements, 
 operators $Q:\ \mathcal{K}(\mathcal{H}^*)\rightarrow\mathcal{K}^(\mathcal{H})$  are strongly convergent series
\begin{equation}                                                                                            \label{eq:hyper}
 e^{\widehat{z}}e^{\widehat{\overline{z}}}\ =\ \sum_{m,n=1}^\infty \frac{\widehat{z}^m\widehat{\overline{z}}^n}{m!n!},\quad
 e^{\widehat{z}+\widehat{\overline{z}}}\ =\ \sum_{n=1}^\infty \frac{(\widehat{z}+\widehat{\overline{z}})^n}{n!},\quad  e^{\widehat{\overline{z}}}e^{\widehat{z}}\ =\ \sum_{m,n=1}^\infty \frac{\widehat{z}^n\widehat{\overline{z}}^m}{m!n!}. 
\end{equation}
Thus any   operator $Q:\ \mathcal{K}(\mathcal{H}^*)\rightarrow\mathcal{K}^(\mathcal{H})$ is a  \emph{partial differential  operator of infinite order with holomorphic coefficients} (see \textsc{\small berezin}\cite[Chapter 1]{Berezin-65}). 

 If a symbol $\sigma^Q(\overline{z},z)$ is a sesqui-holomorhic polynomial 
 (i. e. $\Theta^Q(\overline{z_1},z_2)$ is polynomial on finite-dimensional  sesqui-holomorhic  subspaces
 of $\overline{\mathcal{H}}\times\mathcal{H}$) then $Q$ is a differential operator of finite order and therefore
a   continuous linear operator in $\mathcal{K}(\mathcal{H}^*)$.

\subsubsection{Number operator} 
Let  $\{e_j\}\subset\mathcal{H}$ be an  ortnonormal basis in $\mathcal{H}^0$, so that any $z\in\mathcal{H}$
has a unique orthogonal expansion   $z=\sum_jz_j$ in $\mathcal{H}^0$. This defines a continuous linear\emph{Number operator} in $\mathcal{H}$ 
\begin{equation}
\mathbf{N}\ :=\ \sum_j\widehat{\overline{z_j}}\widehat{z_j}: \mathcal{H}\rightarrow \mathcal{H}
\end{equation}
that  does not depend on orthonormal basis choice.

Furthermore the number operator is non-negative the domain $\mathcal{H}$.
Its Friedrichs Hermitian  extension (again denoted $\mathbf{N}$) has an eigen-basis of monomial functionals $\prod_{k=1}^n(z^*z_{j_k})$ with eigenvalues $n=0,1,...$. Thus the spectrum of 
$\mathbf{N}$ is the point one, the fundamental eigenvalue $0$ is simple but all others are infinitely degenerate.

It follows that the diagonal  matrix elements $\langle e^z|\mathbf{N}|e^z\rangle=(\overline{z}z)e^{\overline{z}z}$ so that the symbols  of $\mathbf{N}$ are
\begin{equation}
\label{eq:symbN}
\sigma_\nu^\mathbf{N}\ \stackrel{(\ref{eq:n2})}{=}\ \overline{z}z,\quad
\sigma_\omega^\mathbf{N}\ \stackrel{(\ref{eq:w})}{=}\ \overline{z}z\ -\ 1/2,\quad
\sigma_\alpha^\mathbf{N}\ \stackrel{(\ref{eq:a})}{=}\ \overline{z}z\ -\ 1.
\end{equation}
\subsection{Finite-dimensional case}
A \emph{Fock representation of canonical commutation relations} over a 
pre-Hilbert space $\mathcal{H}$  is a set  of  linear annihilation and creation operators 
$\widehat{\overline{z}},\ \widehat{z},\ , z\in\mathcal{H})$  in a complex Hilbert space 
$\mathcal{K}(\mathcal{H}^0)$  defined on a  dense subspace
 $\mathcal{K}(\mathcal{H})$ such that
\begin{itemize}
\item $\widehat{\overline{z}},\ \widehat{z}:\mathcal{K}(\mathcal{H})\rightarrow\mathcal{K}(\mathcal{H})$.
\item  The non-zero canonical commutators  relations
\begin{equation}
[\widehat{\overline{z}},\widehat{w}]\ =\ \overline{z}w.
\end{equation}
are satisfied.
\item There is a unit  fiducial $\Omega\in\mathcal{K}(\mathcal{H})$ such that $\mathcal{K}(\mathcal{H})$ is the  linear span of the monomial vectors $\widehat{z}^k\Omega,\ k=0,1,...,$.
\item $\widehat{\overline{z}}\Omega=0$. 
\end{itemize}
As well known (see \textsc{\small glimm-jaffe}\cite[Theorem 6.3.4.]{Glimm87})  for a given $\mathcal{H}$, the Fock representations are irreducible and unitarily equivalent. Furthermore unitary operators defining the equivalence are  completely   defined by the  correspondence between the $\Omega$'s.

\begin{remark} 
Since the sesqui-holomorphic representation is a Fock representation, any
Fock representation sets up the corresponding  nuclear Gelfand  sesqui-linear rigging $\mathcal{K}(\mathcal{H})\subset\mathcal{K}(\mathcal{H}^0)\subset\mathcal{K}(\mathcal{H}^*)$ over the nuclear Gelfand  sesqui-linear rigging $\mathcal{H}\subset\mathcal{H}^0\subset\mathcal{H}^*$. Thus the theory of  partial differential  operators is transferred to all Fock representations. By the unitary equivalence, the operator  calculus formulas are the same (though the unitarily equivalent  realizations of the operators  depend on a representation).
\end{remark}
For a Gelfand  sesqui-linear rigging $\mathcal{H}\subset\mathcal{H}^0\subset\mathcal{H}^*$, the real parts  $x:=(1/\sqrt{2}(z^*+z))$  of  $z\in\mathcal{H}^0$ define  the real Hilbert spaces $X^0:=\Re\,\mathcal{H}^0$ and then the Gaussian real nuclear Gelfand  triple
\begin{equation}
X\ \subset\ X^0\ \subset\ X'.
\end{equation}
By \textsc{\small glimm-jaffe}\cite[Theorem 6.3.3]{Glimm87}), the creation 
annihilation operators
\begin{equation}
\widehat{z}\:\Psi(w'):=\big(x\cdot  w'-\partial_x\big)\Psi(w'),\quad \widehat{z^*}\Psi(w')=\partial_x\Psi(w')
\end{equation}
 together with $\Omega(x)=1$
define the irreducible Fock representation of the commutation relations in the complex Hilbert space
$\mathcal{K}(\mathcal{H}^0)\ :=\ \mathcal{L}^2(\mathcal{H}^*,\ d\gamma$. Then the unitary
equivalence with the representation of the commutation relations in $\mathcal{K}(\mathcal{H}^0)
:=\mathcal{L}^2(\mathcal{H}^*,\ d\gamma)$ produces the equivalent corresponding Gelfand nuclear triple 
$\mathcal{K} \subset\ \mathcal{K}(\mathcal{H}^0)\ \subset\ \mathcal{K}(\mathcal{H}^*)$.

By the fundamental von Neumann-Stone theorem,  all irreducible representations of canonical commutation relations  on a finite-dimensional 
$\mathcal{H}=\mathbb{C}^n$ are unitary equivalent, in particular to Fock representations. 

The Schr\"{o}dinger irreducible representation
is defined by   the Hermitian  unbounded operators $\hat{x},\ x\in\mathbb{R}^n=\Re\mathbb{C}^n$  of the directional multiplications and the operators $\hat{y},\ y\in\mathbb{R}^n=\Im\mathbb{C}^n,$ of the directional  derivatives in the complex Hilbert space  $\mathcal{L}^2(\mathbb{R}^n,d w)$:
\begin{equation}
\hat{x}f(w):=(x\cdot  w)f(w),\quad \hat{y}f(w):=-i(\partial_e\cdot w)f(w).
\end{equation}
The non-zero commutation relation
\begin{equation}
[\hat{y},\hat{x}]=-i(y\cdot x)1 
\end{equation}
leads  to the  creation and annihilation operators $\widehat{z}$ and $\widehat{z^*}$ 
for $z:=(1/\sqrt{2})(x+iy),\ z^*:=1/\sqrt{2})(x-iy)$ in 
$\mathcal{L}^2(\mathbb{R}^n,dx)$ are 
\begin{equation}                                                                                            
\widehat{z}\:\ :\ =(1/\sqrt{2})(\hat{x}+i\hat{y}), \quad\widehat{z^*}\ :\ =(1/\sqrt{2})(\hat{x}-i\hat{y})
\end{equation}
One may choose a Schr\"{o}dinger unit fiducial state as $\Omega(w):=(2\pi)^{-n/2}e^{- w\cdot  w/2}$.

The  fiducial transformation $(2\pi)^{-n/2}e^{- w\cdot  w/2}\mapsto 1$ extends to the unitary isomorphism between $\mathcal{L}^2(\mathbf{R}^n,\ d\gamma)$ and $\mathcal{L}^2(\mathbf{R}^n,\ 
d\gamma)$.
\begin{lemma}
\label{pr:Berezin}
If anti-normal symbol $\sigma^Q_\alpha\ \geq\ 0$  then $Q$ is a non-negatve operator.
 \end{lemma}
\proof It suffices to prove the Lemma in the cylindrical formalism.  There  this is equivalent   to \textsc{\small shubin}\cite[Proposition 24.1]{Shubin} for 
 Schr\"{o}dinger representation of operators on $\mathbf{R}^n$.  \qed 
\begin{lemma}
\label{pr:Folland}
If Weyl symbol $\sigma^Q_\omega=f(\Re z)$  does not depend on $\overline{z}$) then $Q$ is the operator of multilpication with $f\in\mathcal{K}(\mathcal{H})$.
\end{lemma}
\proof It suffices to prove the Lemma in the cylindrical formalism.  There  this is equivalent   to \textsc{\small folland}\cite[Proposition (2.8)]{Folland} for 
 Schr\"{o}dinger representation of operators on $\mathbf{R}^n$.  \qed 

\textsc{\small  agarwal-wolf}\cite{Agarwal} have developed  a comprehensive    calculus of operators of creation and annihilation in finite dimension.  It is quite straightforward to make it mathematically rigorous and then to extend   to quantum field theory via cylindrical approximations.  This would be  another way to deduce results of this section \footnote{ Their formulas are somewhat different because they used symplectic Fourier transform on the phase space. However the  translation of  formulas to the language  of sesqui-holomorphic Borel transform is straightforward (see \textsc{\small folland}\cite[Page 7]{Folland}).}

\subsection{Cylindrical formalism}

A Hermitian  finite-dimensional projector 
$p:\ \mathcal{H}^*\rightarrow\mathcal{H}$ of a rank $n$  induces the \emph{cylindrical  projector}
\begin{equation*}                                                                                           
P\Phi(z^*)\  :=\ \Phi(pz^*), \quad P\Psi(w)\  :=\ \Psi(pw).
\end{equation*}
in $\mathcal{K} \mathcal{H}^*$ and  $\mathcal{K} \mathcal{H}$.

The range of a rank $n$ projector $p$ is naturally isomorphic to $\mathbf{C}^n$. Therefore the cylindrical nuclear Gelfand triples $P(\mathcal{K}(\mathcal{H}^*)\subset P\mathcal{K}(\mathcal{H}^0)\subset P\mathcal{K}(\mathcal{H})$ are equivalent to the Kree triples  over the  finite-dimensional  triples  $\overline{\mathbf{C}}^n\subset\mathbf{C}^n \simeq\overline{\mathbf{C}}^n\subset\mathbf{C}^n$ (see \textsc{\small kree-raczka}\cite{Kree78}). 
  
 The   compressions   of a continuous linear operator $Q:\mathcal{K}(\mathcal{H}^*)\rightarrow
 \mathcal{K}(\mathcal{H})$ are  \emph{cylindrical}   operators  $PQP:P\mathcal{K}(\mathcal{H}^*)\rightarrow
 P\mathcal{K}(\mathcal{H})$. Their  coherent  matrix elements,
\begin{equation}                                                                                            
\label{eq:compression}
 \langle e^{z}\ |\ PQP\ |\ e^z  \rangle\ =\ \langle\ e^{pz}\ |\ Q\ | \ e^{pz}\ \rangle,
 \end{equation}
define    continuous linear operators  from $\mathcal{K}(\mathbf{C}^n)$ to
$\mathcal{K}(\overline{\mathbf{C}}^n)$, i.e.  partial differential operators  of infinite  order on $\mathbf{C}^n$.
\begin{theorem}
\label{pr:cylindrical}
 Operator $Q$ is the strong limit of the cylindrical  differential operators $PQP$  as Hermitian  projectors $p$
converge strongly to the unit operator in $\mathcal{K}(\mathcal{H}^*)$.
\end{theorem}
\textsc{\small  proof}\ 
The coherent matrix elements  $\langle\Psi^{*}|Q|\Phi\ \rangle$ are  separately continuous sesqui-linear forms on  the Frechet space $\mathcal{K}(\mathcal{H}^*)$. By a Banach theorem (see  \cite[Theorem V.7]{Reed1}), the sesqui-linear form is actually continuous 
on $\mathcal{K}(\mathcal{H})$. In particular  operator $Q$ is the weak limit of $PQP$ in $\mathcal{K}(\mathcal{H})$. The nuclear space  $\mathcal{K}(\mathcal{H}^*)$ is a Montel space (see \textsc{\small treves}
\cite[Proposition 50.2]{Treves}). Hence the  weak convergence implies  the strong one in  the topology of $\mathcal{K}(\mathcal{H}^*)$. 
 \qed

As $n\rightarrow \infty$, the matrix elements
 \begin{equation}                                                                                            \label{eq:convergence}
 \langle e^{\overline{z}}\ |\ PQP\ | \ e^w \rangle \ =\ 
\langle\ e^{p\overline{z}}\ |\  Q\ | \ e^{pw}\ \rangle\ 
\rightarrow\  \langle e^{z^*}\ |\  Q\ |   \ e^w \rangle,
\end{equation}
so that coherent matrix elements of the cylindrical $PQP$  converge  to the coherent matrix elements  of $Q$.
 
\section{Classical  Yang-Mills  theory}
\subsection{Yang-Mills fields}
The \emph{global gauge  group}  $\mathbb{G}$ of a  Yang-Mills theory is   a  
connected semisimple compact Lie group with the  Lie algebra $\mathfrak{g}$ of skew-symmetric
 matrices   $X=-X^\prime$. 

The Lie algebra carries the \emph{adjoint representation} $\mbox{Ad}\,(g)X=gXg^{-1}, g\in\mathbb{G}, X\in \mathfrak{g}$, of the group $\mathbb{G}$ and the corresponding selfrepresentation $\mbox{ad}(X)Y=[X,Y],\ X,Y\in\mathfrak{g}$.  The adjoint  representation is orthogonal with respect to the \emph{positive  definite}  Ad-invariant  scalar  product
\begin{equation}                                                                                            \label{eq:scalar}
X\cdot Y \  :=\ \mbox{trace}(\mbox{ad}X^\prime\mbox{ad}Y)\ =\ -\mbox{trace}(\mbox{ad}X\mbox{ad}Y), 
\end{equation}
the negative Killing form on $\mathfrak{g}$.  

There exists an orthonormal basis $\{T_k\}$ in $\mathfrak{g}$ such that  
\begin{equation}                                                                                            \label{eq:skew}
[T_i,T_j]\ =\ c_{ij}^kT_k,
\end{equation}
with the structure constants $c_{ij}^k$ are skew-symmetric with respect to interchanges of all three  indices $i,j,k$. (Summation over repeated indices is presumed throughout.)
\bigskip
Let the Minkowski space $\mathbb{R}^{1,3}$ be oriented and time oriented with  the Minkowski metric signature 
$(1,-1,-1,-1)$. In a Minkowski coordinate system $x^\mu, \mu=0,1,2,3,$  the metric tensor is diagonal.
  In  the natural unit system, the time coordinate $x^0=t$. Thus  $(x^\mu)=(t,x^i),\  i=1,\ 2,\ 3$. 
  
 The \emph{local gauge  group} $\widetilde{\mathbb{G}}$ is the group of  infinitely differentiable $\mathbb{G}$-valued functions
 $g(x)$ on $\mathbb{R}^{1,3}$ with the pointwise group multiplication.  The  \emph{local gauge Lie algebra}  
 $\tilde{\mathfrak{g}}$ of $\mathfrak{g}$-valued functions   on 
 $\mathbb{R}^{1,3}$ with the pointwise Lie bracket.   
consists of  infinitely differentiable $\mathfrak{g}$-valued functions   on 
 $\mathbb{R}^{1,3}$ with the pointwise Lie bracket.   
 
$\widetilde{\mathbb{G}}$ acts via the pointwise adjoint action on $\widetilde{H})$   and correspondingly on  
$\mathcal{A}$, the real vector space of \emph{gauge   fields}   $A=A_\mu(x)\in\tilde{\mathfrak{g}}$. 

 \smallskip
   Gauge fields $A$ define   the \emph{covariant partial derivatives}  
\begin{equation}
\label{eq:covariant}
  \partial_{A\mu}X\  :=
\  \partial_\mu X- \mbox{ad}(A_\mu)X,\quad
X\in\widetilde{H}.  
\end{equation}
This definition shows that  in the natural units \emph{gauge  connections have the  mass dimension} $1/[L]$. 

Any $\tilde{g}\in\widetilde{\mathbb{G}}$ defines the affine \emph{gauge transformation}  
\begin{equation}                                                                                            \label{}
A_\mu\mapsto A_\mu^{\tilde{g}}:\ =\ \mbox{Ad}\,(\tilde{g})A_\mu-(\partial_\mu \tilde{g})\tilde{g}^{-1},\ A\in \mathcal{A},
\end{equation}
so that $A^{\tilde{g}_1}A^{\tilde{g}_2}=A^{\tilde{g}_1\tilde{g}_2}$.

\medskip
Relativistic Yang-Mills \emph{curvature} $F(A)$ is  the 
antisymmetric tensor \footnote{ The scaleless  Yang-Mills coupling  constant is supressed.}\begin{equation}
\label{eq:curvature}
F(A)_{\mu\nu} :=
\partial_\mu A_\nu-\partial_\nu A_\mu-[A_\mu,A_\nu].
\end{equation} 
The curvature is gauge invariant:
\begin{equation}
 \label{}
\mbox{Ad}\,(g)F(A)\ =\ F(A^{g}), 
 \end{equation}
 as well as \emph{Yang-Mills Lagrangian} 
\begin{equation}
 \label{eq:Lag}
  (1/4)F(A)^{\mu\nu}\cdot F(A)_{\mu\nu}.
  \end{equation}
 The corresponding gauge invariant  Euler-Lagrange equation is a   2nd order non-linear  partial differential equation $\partial_{A\mu}F(A)^{\mu\nu} =0$, called 
 the \emph{Yang-Mills equation} 
\begin{equation}                                                                                           
 \label{eq:YM}
 \partial_\mu F^{\mu\nu}\ -\ [A_\mu, F^{\mu\nu}]\ =\ 0.
\end{equation}
 \emph{Yang-Mills fields} are solutions of Yang-Mills equation.
 
\subsection{Yang-Mills phase space}
 In the temporal gauge  $A_0(t,x^k)=0$  the   2nd order Yang-Mills equation (\ref{eq:YM})  is equivalent to  the 1st order   hyperbolic system   for the time-dependent $A_j(t,x^k)$,  $E_j(t,x^k):=F_{0,x^k}$ on  $\mathbb{B}$ (see   \textsc{\small  goganov-kapitanskii} \cite[Equation (1.3)]{Goganov})
\begin{equation}                                                                                            
\label{eq:evolution}
\partial_t A_k\  = \  E_k,  \quad
\partial_tE_k \  = \  \partial_jF^j_k - [A_j,F^j_k],\ \quad\ F^j_k\ =\ \partial^j A_k - \partial_k A^j - [A^j,A_k].
\end{equation}
and the \emph{constraint  equations}
\begin{equation}                                                                                            
\label{eq:constraint}
 [A^k,E_k]    \ =\  \partial^kE_k, \quad \mbox{i.e.}\ \quad \partial_{k,A}E_k\ =\ 0.
\end{equation}
By  \textsc{\small  goganov-kapitanskii} \cite{Goganov}, the evolution system is  a semilinear first order  partial differential  system  with  \emph{finite propagation speed } of the initial data, and the  initial  problem for it with  constrained initial data at $t=0$ 
\begin{equation}                                                                                           
 \label{eq:initial}
a_k(x)\  :=\ A(0,x_k), \ e_k(x)\  :=\ E(0,x_k), \quad \partial^ke_k=[a_k,e,_k]
\end{equation}
is \emph{globally and uniquely solvable} in local Sobolev spaces on the whole Minkowski space  $\mathbb{R}^{1,3}$ (with no restrictions at the space infinity.)

This fundamental theorem has been  derived via  Ladyzhenskaya   method (see \cite{Goganov}) by a reduction to initial data with compact supports.
If the  constraint equations are satisfied  at $t=0$, then, by (\ref{eq:evolution}, they are satisfied   for  all $t$ automatically. Thus the  \emph{1st order evolution system along with the  constraint equations for initial data is equivalent  to the 2nd order Yang-Mills system}. Moreover the constraint equations are invariant under  \emph{time independent} gauge transformations.

 Sobolev-Hilbert spaces $\mathcal{A}^s, -\infty<s<\infty,$ of (generalized) connections $a(x)$ are the  completions of of the spaces of smooth connections with compact supports in open balls    $\mathbb{B}$ of radius $r$  with respect to the norms 
\begin{equation}                                                                                            \label{eq:SH}
 |a|^2 :=
\int_{\mathbb{B}}\, dx\,\big(a\cdot(1-\triangle)^sa\big) < \infty. 
\end{equation}

They define the real Gelfand nuclear triple (see  \cite{Gelfand}) 
\begin{equation}                                                                                            \label{eq:Gelf}
\mathcal{A}:\mathcal{A}\  :=
 \bigcap\mathcal{A}^s\ \subset \mathcal{A}^0\  \subset\
\mathcal{A}^*  :=
\bigcup\mathcal{A}^s,
\end{equation}
where $\mathcal{A}$ is  a \emph{real} nuclear Frechet  space with the dual $A^*$. 

Similarly we define the chain of Sobolev-Hilbert spaces $\mathcal{S}^s, -\infty<s<\infty,$ of (generalized)  scalar fields $u(x)$  on  $\mathbb{B}$ with values in $\mbox{Ad}\:\mathbb{G}$ and the Hilbert norms $|u|$. Let 
\begin{equation}                                                                                            \label{eq:GelfS}
\mathcal{S}:\mathcal{S}\  :=
 \bigcap\mathcal{S}^s\ \subset \mathcal{S}^0\  \subset\
\mathcal{S}^*  :=
\bigcup\mathcal{S}^s
\end{equation}
be the corresponding Gelfand  rigging.
 
Let   $a\in\mathcal{A}^{s+3},\ s\geq 0$. Then, by Sobolev embedding theorem $a$  is  continuously $s+2$-differentiable  on  $\mathbb{B}$ and, therefore, the following  gauged  vector calculus  operators are continuous:
\begin{itemize}
  \item \emph{Gauged gradient} $\mbox{grad}^{[a]} \ :\  \mathcal{S}^{s+1}\rightarrow  \mathcal{A}^s$,
 \begin{equation}                                                                                            \label{} 
\mbox{grad}^{[a]}_{k}u \  :=\  \partial_k u - [a_k,u].
\end{equation}

 \item \emph{Gauged divergence} $\mbox{div}^{[a]}\ :\  \mathcal{A}^{s+1}\rightarrow  \mathcal{S}^s$,
 \begin{equation}                                                                                            \label{eq:div}  
\mbox{div}^{[a]}\:b \  := \   \mbox{div}\:b - [ a\ ;\ b], \quad  [ a\ ;\ b]:= [a_k, b_k].
\end{equation}
 \item \emph{Gauged curl} $\mbox{curl}^{[a]}\  :\ \mathcal{A}^{s+1}\rightarrow  \mathcal{A}^s$,
\begin{equation}
 \label{eq:curl}  
 \mbox{curl}^{[a]}b \  :=\ \mbox{curl}\:b - [a\stackrel{\times}{,}b], \quad  [a\stackrel{\times}{,}b]_i\  :=\ \varepsilon_{ij}^k\ [a_j,b_k]. 
 \end{equation}
  \item \emph{Gauged Laplacian}  $\triangle^{[a]}:\ \mathcal{S}^{s+2}\rightarrow \mathcal{S}^s$,
 \begin{equation}
  \label{eq:Laplacian}  
\triangle^{[a]} u\  :=\ \mbox{div}^{[a]}(\mbox{grad}^{[a]}u).
  \end{equation}
\end{itemize} 
 The  adjoints of the gauged   operators  are
\begin{equation}                                                                                            \label{eq:adjoint}
 \mbox{grad}^{a*}  =\ -\mbox{div}^{[a]}.
\end{equation}
\begin{lemma}
\label{pr:sur}
If  $a\in\mathcal{A}^{s+3}, s\geq 0,$ then  the  operator  $\mbox{div}^{[a]}:\mathcal{A}^{s+1}\rightarrow \mathcal{A}^s$ is surjective.
\end{lemma}
\proof
Let $\mathring{\mathcal{S}}^{s+2},\ s\geq 0,$  denote  the closure in $\mathcal{S}^{s+2}$ of the space of $u$'s with compact support in the interior of $\mathbb{B}$.
The conventional Laplacian  $\triangle^0:\ \mathring{\mathcal{S}}^{s+2}\rightarrow \mathcal{S}^s$ is an isomorphism (see  \textsc{\small  agranovich}\cite{Agranovich}).

The gauged Laplacian $\triangle^{[a]}$   differs from the usual  Laplacian $\triangle^0$ by first order differential operators, and, therefore is a Fredholm operator of zero  index from $\mathring{\mathcal{S}}^{s+2}$ to  $\mathcal{S}^s,\ s\geq 0$.

If $\triangle^{[a]}u=0$ then then  the $*$-product $(\triangle^{[a]}u)^* u =  (\mbox{grad}^{[a]}\,u)^* (\mbox{grad}^{[a]}\,u)$  so that  $\mbox{grad}\,u=[a,u]$.
The computation
\begin{equation}                                                                                            \label{}
(1/2)\partial_k(u\cdot u)= (\partial_ku\cdot u)=[a_k,u]\cdot u= -\mbox{trace}(a_kuu-ua_ku)=0
\end{equation}
shows that  the solutions    $u\in\mathring{\mathcal{S}}^{s+2}$ are constant. Because they vanish on the ball boundary, they vanish on the whole ball.   Since the index of the Fredholm operator $\triangle^{[a]}$ is zero,  its range  is a closed subspace with the codimension equal to the dimension  of its  null space. Thus    the operator $\mbox{div}^{[a]}\mbox{grad}^{[a]}$  is surjective, and so is $\mbox{div}^{[a]}$. \qed

\medskip
 Consider the  bundles $\mathcal{H}^s, s\geq 0,$ of constraint initial data   with the base  
$\mathcal{A}$ and the null vector spaces $\mathcal{E}^{s+1}_a$ of the operators $\mbox{div}^{[a]}:\ \mathcal{E}^{s+1}\rightarrow \mathcal{E}^{s}$ as fibers over  $a\in\mathcal{A}$. 
  
   Their intersection $\mathcal{H}$ is  a vector bundle of nuclear countably Hilbert spaces over the nuclear countably Hilbert base   $\mathcal{A}$.  Together with the unions of the dual spaces $\mathcal{H}^{-s}$ they form  a bundle of nuclear Gelfand triples $\mathcal{H}$ over the same base.
\begin{proposition}
\label{pr:orthogonal}
The  bundle  $\mathcal{H}$ is  smoothly \footnote{ In this paper, smooth = infinitely differentiable.} trivial,  so that the total space of  $\mathcal{H}$ is smoothly isomorphic to the Hilbert direct product  of  its base  
$\mathcal{A}$ and the transversal  fiber  $\mathcal{E}_\bot:=\mbox{div}(e)=0$ over $a=0$.
\end{proposition}
\proof

The equation for $(a,e)\in\mathcal{H}\times\mathcal{H}$
\begin{equation}
\label{eq:implicit}
\mbox{div}^{[a]}e-\mbox{div}^{[0]}e\ =\ 0, \quad \mbox{div}^{[0]}e=\mbox{div}e,
\end{equation}
 is satisfied  on the vector space  $(0,e):\ =\ \mbox{div}e=0$.  
 
 Furthermore,  the mapping $\phi(a,e):\mbox{div}^{[a]}e-\mbox{div}^0e$ uniformly satisfies   the conditions of Nash-Moser implicit function theorem in the form of  \textsc{\small raymond}\cite{Raymond} on a neighborhood  of $\mathcal{E}_\bot$ (the only non-routine  condition of  surjectivity of the partial Frechet derivative $\partial_e\phi(a,e)$ is  provided by  Lemma \ref{pr:sur}). Hence there exists  a smooth explicit mapping $e=e(h)$ on that neighborhood such that  $(a,e(a))$ solves the equation (\ref{eq:implicit}).  
 
The mappings $e=e(h)$  are smooth local  trivializations  of the vector  bundle $\mathcal{H}$.  It 
is associated with the locally  trivial bundle of smooth $*$-orthonormal frames in the fibers (with respect to
the scalar product in  $\mathcal{E}_\bot^0$).

Since the Frechet space $\mathcal{A}$ is paracompact, its smooth homothety retraction to the origin $a=0$ has a homotopy lifting to the frames space (see \textsc{\small nash-sen}\cite[Section 7.6]{Nash}). Thus the bundle $\mathcal{H}$ is trivial, so that the total set of constraint initial data is converted into the  Hilbert space  $\mathcal{A} \times \ \mathcal{H}_{a=0}$ with the flat parallel transport preserving $e\cdot e$.\qed

\subsection{Classical Yang-Mills Hamiltonian}

The \emph{scaleless Yang-Mills Hamiltonian} with the concealed coupling constant is (see {\small  \textsc{\small glassey-strauss}\cite[Equation (10)] {Glassey} ) 
 \begin{equation}                                                                                            
 \label{eq:N1}
H(a,e):=\  (r/2)\int_{\mathbb{B}(r)}\,d^3x\:
\big((\mbox{curl}\,a -  [a\stackrel{\times}{,}a])\cdot (\mbox{curl}\,a -  
[a\stackrel{\times}{,}a]) + e\cdot e \big),\quad (a,e)\in\mathcal{A}\times\mathcal{E},  
\end{equation}
where $[a\stackrel{\times}{,}a]$ is the vector field with the $i$-th  components $\varepsilon_{ij}^k[a_j,a_k]$ is invariant with respect to scaling $x\rightarrow sx,\ x\in \mathbf{R}^3, 0<s<\infty$. (The factor $r$ makes   the functional  to be scaleless.)

The integrand  of 
\begin{equation}                                                                                            
 \label{eq:N2}
H_1(a):=\  (r/2)\int_{\mathbb{B}(r)}\,d^3x\:
\big((\mbox{curl}\,a -  [a\stackrel{\times}{,}a])\cdot (\mbox{curl}\,a -  [a\stackrel{\times}{,}a]\big),
\end{equation}
 is the curvature of  the time-independent  gauge fields $a(x)$. Thus  $H(a)$ is invariant under time-independent gauge transformations and therefore is constant on each orbit of the smooth local 
 time-independent gauge group.

 By \textsc{\small dell'antoniio-zwanziger}\cite[Proposition 1]{Dell'Antonio}, the closure of the local gauge Lie
group  $\widetilde{\mathbb{G}}^1$ in the Sobolev space $\mathcal{A}^1$ is an infinite-dimenssional compact group with a continuous action in the Hilbert space  $\mathcal{A}^0$. The  action orbits are compact so that 
the squared continuous Hilbert  norm $\|a\|^2$ has an  absolute minimal value on every orbit. The minimal values are attained at transversal $a:\ \mbox{div}\:a=0$ (in the distributional  sense).

By Sobolev embedding theorem, $\mathcal{A}^1\subset\mathcal{L}^6(\mathbb{B})$. Therefore the functional
$H_1(a)$ has a unique continuation to  $\widetilde{\mathbb{G}}^1$-orbits in $\mathcal{A}^0$ and is constant on each of them.

All in all, the following proposition holds: 
\begin{proposition}
The constrained Yang-Mills Hamiltonian $H(a,e)$ is completely determined by its restriction to the countably Hilbert 
vector space $\mathcal{H}:\ \mathcal{A}_\bot\times \mathcal{E}_\bot$ of transversal  constrained vector fields $(a,e)$.
\end{proposition}

 The complex Yang-Mills fields
\begin{equation}
\label{eq:dimension}
z\ :=\ (r/\sqrt{2})(a+ire),\quad  \overline{z}\ := \ (r/\sqrt{2})(a-ire)
\end{equation}
are scaleless, as well as their    Hermitian  product 
\begin{equation}
\label{eq:prod}
\overline{z}z\ :=\ (1/r^3)\int_{\mathbb{B}(r)}d^3x\:\overline{z}(x)\cdot z(x).
\end{equation}
Let 
\begin{equation}
\mathcal{Z}\ :=\  \mathcal{A}_\bot+i \mathcal{E}_\bot\ \subset\ 
\mathcal{Z}^0\ :=\  \mathcal{A}_\bot^0+i \mathcal{E}_\bot^0\ \subset\ 
\mathcal{Z}^*\ :=\  \mathcal{A}_\bot^*-i \mathcal{E}_\bot^*
\end{equation}
be the  corresponding scaleless Gelfand triple.

\section{Mathematical  quantum Yang-Mills theory}
\subsection{Quantization of classical Yang-Mills fields}
Yang-Mills equation (\ref{eq:evolution})
is equivalent to  an infinite-dimensional Hamiltonian system for  $a(t):=A(t,x)$ and $e(t))=\partial_tA(t,x)$ on the reduced phase space (see  \textsc{\small faddeev-slavnov}\cite[Chapter III, Equations (2.64)]{Faddeev}) 
\begin{equation}
\label{eq:Ham}
\partial_t a(t)\ =\  -\partial_{e(t)}H(a(t),e(t)),\quad \partial_t e(t)\ =\  \partial_{a(t)}H(a(t),e(t)).
\end{equation}
By the equivalence, solutions exist for all $t$ and are uniquely defined by the initial data $a(0),e(0)$.

Let
\begin{equation}
z(t)\ =\big(a(t)+ie(t)\big)/\sqrt{2}, \quad \overline{z}(t)\ =\big(a(t)-ie(t)\big)/\sqrt{2}.
\end{equation}

By I. Segal quantization (see \textsc{\small reed-simon}\cite[Section X.7]{Reed2}), the compactly supported classical fields $a(t)$ and $e(t)$ in $\mathbb{B}$
are  quantized as the symmetric  operators in  $\mathcal{K}(\mathcal{H}^*)$
\begin{equation}
\label{eq:quantum}
\widehat{a}(t)\ :=\ (\widehat{z}(t)+\widehat{\overline{z}}(t))/\sqrt{2},\quad 
\widehat{e}(t)\ :=\ -i(\widehat{z}(t)-\widehat{\overline{z}}(t))/\sqrt{2}
\end{equation}
 understood as  local  quantum fields on Minkowski space.

\subsubsection{Adapted G\r{a}rding-Wightman axioms} 
\begin{itemize}
\item For a proof of essential  self-adjointness of $\widehat{a}(t),-i\widehat{e}(t)$, irreducibility of the set of all local quantum Yang-Mills fields, uniqueness (up to phase transformations) of the vacuum Fock state, and bosonic canonical commutations relations see \textsc{\small reed-simon}\cite[Theorem X.41]{Reed2}.   The corresponding Hermitian extensions in  $\mathcal{K}(\mathcal{H}^0)$ are denoted $\widehat{a}(t),-i\widehat{e}(t)$ again. 
\item
Each of  normal, Weyl, and anti-normal symbols of the operator $\widehat{a}(t)$ is  $\overline{z}(t)+z(t)$ 
  satisfies  Yang-Mills equations on Minkowski space. Therefore,   the local Yang-Mills quantum fields do the same. 
\item
Poincar\'{e} covariance of local Yang-Mills quantum fields follows from the covariance of their symbol.
\item
Similarly,  the evolution of the symbol has the  finite  propagation speed property, and so does  tne evolution of quantum Yang-Mills fields, as required  by Einstein causality principle.

Furthermore, if the compact supports of $z(t_1)$ and $z(t_2)$  on $\mathbb{R}^3$ are disjoint then the commutators 
\begin{equation}
[\widehat{a}(t_1),\widehat{a}(t_2)]\ =\  \big[\widehat{z}(t_1)+\widehat{\overline{z}}(t_1)/\sqrt{2},\ 
\widehat{z}(t_2)+\widehat{\overline{z}}(t_2)/\sqrt{2}\big]\ =\ i\ \Im (\overline{z}(t_1)z(t_2))\ =\ 0.
\end{equation}

\item
The spectral  positivity  of  quantum Yang-Mills Hamiltonian  follows  from the non-negativity of its  anti-normal symbol. 
\end{itemize}

\subsection{Symbols of quantum Yang-Mills Hamiltonian}

Since Yang-Mills Hamiltonian $H=rP^0$ is scaling invariant, it is sufficient to take $r=1$.

The  quantum quantum Yang-Mills Hamiltonian is the infinite-dimensional   partial differential operator $\widehat{H}:\mathcal{K}(\mathcal{H}^*)\rightarrow\mathcal{K}(\mathcal{H})$
with   the classical Yang-Mills  Hamiltonian (\ref{eq:N1}, $r=1$) as the anti-normal symbol
\begin{equation} 
\label{eq:Schr}    
\sigma_\alpha^{\widehat{H}}:=H(\overline{z},z)=H(a,e)=:H_1(a)+H_2(e),
\end{equation}
The following Proposition is crucial (note that $ a=(1/\sqrt{2})(\overline{z}+z),\ e=(i/\sqrt{2})(\overline{z}-z)$.
\begin{proposition}
\label{pr:crucial}
The Weyl and normal  symbols  of the anti-normal quantum Yang-Mills Hamiltonian $\widehat{H}$ are equal to
\begin{eqnarray}
& &                                                                                            
\label{eq:crucial}
\sigma^{\widehat{H}}_\omega\ =\ H_1(a)\ +\  (1/2)\|a\|^2\ +\ \overline{z}z\ +\ 9/16,\\   
& &
\label{eq:nucrucial}
\sigma^{\widehat{H}}_\nu\ =\ H_(a,e)\  + \ \|a\|^2\  +\ 24/16.
\end{eqnarray}
\end{proposition}
\proof
In view of  cylindrical approximations, it suffices  to verify Equation (\ref{eq:crucial})  in the Schr\"{o}dinger representation of the canonical commutation relations over finite-dimensional Euclidean spaces.

The operator  $\partial_{\overline{z}*}\partial_z=(1/2)(\Delta_a+\Delta_e)$, where the Laplacians are defined with respect to the Killing form on the gauge Lie algebra $\mathfrak{g}$.

By  (\ref{eq:omeganu}),  
\begin{equation}                                                                                           
 \label{eq:W}
\sigma^{\widehat{H}}_\omega(a,e)\ =\
\Big(1\ +\ \frac{\Delta_a/2}{2}\ +\ \frac{(\Delta_a/2)^2}{2}\Big)H_1(a)\ +\  \Big(1+\frac{\Delta_e/2}2\Big)H_2(e).
\end{equation}
Differentiation with respect to $x$ is a continuous linear operator in $\mathcal{H}^*$. It acts naturally in
$\mathcal{K}(\mathcal{H}^*)$   commuting    with  $\Delta_a$ and $\Delta_e$. Hence, by   Leibniz rule for diifferentiation with respect to $a$,  one has 
\begin{eqnarray}
& &
\nonumber
 \Delta_a(\mbox{curl}\,a\cdot\mbox{curl}\,a)\\
& &
\nonumber
 =\ \Delta_a(\mbox{curl}\,a)\cdot\mbox{curl}\,a +
2\nabla_a\mbox{curl}\,a \cdot\nabla_a\mbox{curl}\,a+ \mbox{curl}\,a\cdot \Delta_a\mbox{curl}\,a\\
& &
\label{eq:N31}
=\ (\mbox{curl}\,\Delta_aa)\cdot\mbox{curl}\,a +
2\mbox{curl}\,\nabla_aa \cdot\nabla_a\mbox{curl}\,a+ \mbox{curl}\,a\cdot \mbox{curl}\,\Delta_aa\ =\ 0
\end{eqnarray}
where in the third line the partial differentions with respect to the variables $x$ and $a$ are interchanged 
and  $\Delta_aa=0$ and $\nabla_a a$ is a constant matrix field.

Similarly,
\begin{equation}                                                                                            
\label{eq:N32}
  \Delta_a\big(\mbox{curl}\,a\cdot[a\stackrel{\times}{,}a]\big)
=\mbox{curl}\,\Delta_aa\cdot[a\stackrel{\times}{,}a]+
2(\mbox{curl}\,\nabla_a a)\cdot\nabla_a [a\stackrel{\times}{,}a]+ 
a\cdot \mbox{curl}\, \Delta_a[a\stackrel{\times}{,}a]\ =\ 0,
\end{equation}
since $\Delta_aa=0$, curl is a symmetric differential  operator. and $\nabla_a[a\stackrel{\times}{,}a]$ is a constant matrix field.

\medskip
Next, in  the   orthonormal basis $T_k$ (\ref{eq:skew}) of $\mathfrak{g}$   the  structure constants $c_{ij}^k$ of the semisimple Lie algebra are totally anti-symmetric. Thus
\begin{equation}
 [a_i,a_j]=\sum_{k}c_{ijk}a_i^ka_j^k,\quad a_i=a_i^kT_k.
\end{equation}
Then, by \textsc{\small simon}\cite[page 217]{Simon},
\begin{equation}                                                                                            
\label{eq:Simon}
[a_i,a_j]\cdot[a_i,a_j]\ =\  \sum_k(a^k_ia^k_jc_{ij}^k)^2.
\end{equation}

In the cartesian coordinates $\alpha^k_i(x)$, the differential operator  $\Delta_a$  is the standard  Laplacian with respect to $a$ independently of $x$ so that, by \textsc{\small  simon}\cite[page 217]{Simon}),
 \begin{equation}                                                                                            \label{eq:0}
\Delta_aH_1(a)\ =\ \int_{\mathbb{B}}\,dx\:(1/2)\partial^2/\partial(\alpha^k_i(x))^2
2\sum_k(\alpha^k_i\alpha^k_jc_{ijk })^2(x).
\end{equation}
The sqew-symmetry of $c_{ij}^k$ implies that  $\sum_k\alpha^k_i\alpha^k_jc_{ij}^k$  does not contain  
$(\alpha^k_i)^2$. Then, by the Leibniz rule, 
\begin{eqnarray*}
& &
\partial^2/\partial(\alpha^k_i)^2\sum_k(\alpha^k_i\alpha^k_jc_{ijk })^2\ =\
2\big(\partial^2/\partial(\alpha^k_i)^2\sum_k(\alpha^k_i\alpha^k_jc_{ijk }\big)
(\sum_k(\alpha^k_i\alpha^k_jc_{ijk }\big)\\
& &
+\ 2\big(\partial/\partial\alpha^k_i\sum_k\alpha^k_i\alpha^k_jc_{ij}^k\big)
\big(\partial/\partial\alpha^k_i\sum_k\alpha^k_i\alpha^k_jc_{ij}^k\big)\\
& &
=\ 2\sum_{ijkl}\alpha^k_ic_{ij}^k\alpha^l_jc_{ljk}(x)\ =\ 2a(x)\cdot a(x)\quad 
\mbox{(see  \textsc{\small  simon}\cite[page 217]{Simon})}.
\end{eqnarray*} 
Thus
\begin{equation}                                                                                            
\label{eq:1}
\Delta_aH_1(a)\ =\   \int_{\mathbb{B}}\, dx\,a\cdot a\ =\|a\|^2,\quad \Delta_a^2H_1(a)=\Delta_a\|a\|^2=2
\end{equation}
Furthermore, 
\begin{equation}                                                                                            
\label{eq:2}
\Delta_eH_2(e)\ =\ \Delta_e\|e\|^2\ =\ 2.
\end{equation}
 Equations   (\ref{eq:W}), (\ref{eq:N31}),  (\ref{eq:N32}), (\ref{eq:0}), (\ref{eq:1}), (\ref{eq:2}) entail Equation (\ref{eq:crucial}) of the Proposition for the Weyl symbol $\sigma_\omega^{\widehat{H}}$.
 
Equation (\ref{eq:nucrucial}) for the normal symbol $\sigma_\omega^{\widehat{H}}$ is derived in the same way but, in view of (\ref{eq:omeganu}), with  $(1/2)\Delta_a,\ (1/2)\Delta_e$ in (\ref{eq:W}) replaced with 
$\Delta_a,\ \Delta_e$. 
 \qed   
 
\subsection{Spectrum of quantum Yang-Mills Hamiltonian}
The anti-normal symbol of quantum Yang-Mills Hamiltonian  $\widehat{H}$
is non-negative so that, by Lemma \ref{pr:Berezin},  $\widehat{H}$ is non-negative  and, in particular, symmetric on the dense domain $\mathcal{H}$ in  the Hilbert space $\mathcal{H}^0$. As such it has the Friedrichs Hermitian  extension $\widehat{H}$ (by abuse of the notation, it is transfered to the Hermitian  operator). Now the spectrum of  $\widehat{H}$ is uniquely defined.

By definition, a  \emph{positive  mass  gap} means that the lowest eigenvalue is simple and an isolated spectral value corresponding to the physical vacuum. Then the the lowest boundary value of the rest of the spectrum represents  the \emph{ physical mass}.
\begin{theorem}
\label{pr:mass}  
 Yang-Mills Schroedinger spectrum has a positive mass gap $\geq 9/16$.
 \end{theorem}
\proof
The lowest spectral value of $\widehat{H}$
\begin{equation}
\label{eq:fundamental}
\lambda_1(\widehat{H})\ \leq\ \inf \sigma_\nu^{\widehat{H}}(\overline{z},z)\ \stackrel{(\ref{eq:Schr})}{=}\ 24/16,
\end{equation}
because, by (\ref{eq:n2}), the normal symbol $\sigma_\nu^{\widehat{H}}(\overline{z},z)$ is the expectation  value of the operator $\widehat{H}$ on the coherent states. 

On the other hand, by Proposition \ref{pr:Folland}, $H_1(a)\ +\ \|a\|^2/2$ is Weyl symbol of the multiplication operator   with $H_1(a)\ +\ \|a\|^2/2\ \geq\ 0$.

Therefore
\begin{equation}
\label{eq:geq}
\sigma_\omega^{\widehat{H}} \ \stackrel{(\ref{eq:symbN})}{=}\  (\overline{z}z-1/2)+1/2+\ H_1(a)\ +\ \|a\|\ ^2/2\ +\ 9/16\  
 \geq \ \sigma_\omega^{\mathbf{N}}+17/16,
\end{equation}
 where $\mathbf{N}$ is the number operator.

The shifted number operator    $\mathbf{N}+17/16$ has  the simple and isolated fundamental eigenvalue   $17/16$ and no other spectral value in the interval $17/16<\lambda<17/16+1-0$.

 By the inequality  (\ref{eq:fundamental}) and  non-negativity of $\mathbf{N}$,  operator $\widehat{H}$   has a spectral value in the same interval.

Finally, the minimax  variational principle (see \textsc{\small berezin-shubin}\cite[Supplement, Section 3, Corollary 1]{Berezin-91})  implies from (\ref{eq:geq}) that  in the interval $17/16<\lambda<17/16+1-0$  all spectral values of $\widehat{H}+17/16$  are its
eigenvalues and the sum of their multiplicities is not greater than $1$, the sum of  multiplicities of the eigenvalues of $\mathbf{N}+17/16$.

Thus quantum Yang-Mills Hamiltonian has a positive mass gap $\geq (17/16+1)-24/16=9/16$.  \qed

\begin{theorem}
\label{pr:spectrum}
Yang-Mills energy operator has a countable eigenbasis for $\mathcal{F}(\mathcal{H}^0)$ so that its spectrum is a countable set of eigenvalues.
\end{theorem}
\proof

As a complex Hilbert space, the space $\mathcal{H}^0$ is isomorphic to 
$\mathcal{L}_\bot^2(\mathbb{B},\mathbf{C}\mathfrak{g}^3)$ of  transverse square-integrable vector-valued functions $z(x)$ on $\mathbb{B}$ with values in $\mathbf{C}\mathfrak{g}^3$.

 The  Fourier series expansions of $z(x)=\big(z_1(x),z_2(x),z_3(x)\big),\ x\in\mathbb{B},$
\begin{equation}                                                                                           
z_k(x)=\sum_{j_k\in\mathbf{Z}^3}\! \:\check{z}(j_k)\exp(2\pi\: ix\cdot j_k), \quad \check{z}(j_k)\in\mathbf{C}\mathfrak{g}^3, 
\end{equation}
define  the  isomorphism between  $\mathcal{L}_\bot^2(\mathbb{B},\mathbf{C}\mathfrak{g}^3)$ and the Hilbert tensor product $l^2(\mathbf{Z}^3)\otimes
\mathbf{C}\mathfrak{g}^3$ of   square summable
$\mathbf{C}\mathfrak{g}$-valued  sequences (with the natural conjugation), subject to the transversality condition 
\begin{equation}                                                                                           
 \label{eq:transversality}
j_k\cdot \check{z}(j_k)\ =\ 0,\quad j_k\in\mathbf{Z}^3,\ k=1,2,3.
\end{equation}
The isomorphism  converts the Gelfand nuclear triple $\mathcal{H}\subset\mathcal{H}^0\subset\mathcal{H}^*$ into a  Gelfand nuclear triple  of $\mathbf{C}\mathfrak{g}$-valued sequences presented as elements of completed topological tensor products
\begin{equation}                                                                                           
 \label{eq:l}
\ell(\mathbf{Z}^3)\otimes\mathbf{C}\mathfrak{g}\ \subset\ l^2(\mathbf{Z}^3)\otimes\mathbf{C}\mathfrak{g}\ \subset\ \ell^*(\mathbf{Z}^3)\otimes\overline{\mathbf{C}}\mathfrak{g}^*.
\end{equation}
where $\ell(\mathbf{Z}^3)$ is a nuclear Frechet subspace of $l^2(\mathbf{Z}^3)$.

By the infinite-dimensional version of the Hartogs theorem (see \textsc{\small colombeau}\cite[Section 3.3]{Colombeau}),   the corresponding Fock space $\mathcal{K}(\mathcal{H}^*)$ is isomorphic to the space of all continuous functionals on   $\ell^*(\mathbf{Z}^3)\otimes\overline{\mathbf{C}}\mathfrak{g}^*$ that are exponential entire functions  on $\overline{\mathbf{C}}\mathfrak{g}$, and the corresponding Kree space $\mathcal{K}(\mathcal{H}^*)$ is isomorphic to the space of all  functionals on   $\ell(\mathbf{Z}^3)\otimes\mathbf{C}\mathfrak{g}^*$ that are holomorphic functions on  $\mathbf{C}\mathfrak{g}$.

By unitarity of  the Fourier series expansions, the Weyl symbol $\sigma^{\widehat{H}}_\omega$ (\ref{eq:crucial}) is unitarily equivalent  to 
\begin{eqnarray*}
& &
\sigma_\omega^{\widehat{H}_j}(\check{a},\check{e})\ =\  \frac{1}{2}\bigoplus_{j\in\mathbf{Z}^3}\!\Big\|\big(j\times \check{a}(j) - 
[\check{a}(j)\stackrel{\times}{,}\check{a}(j)] \big)\Big\|^2\ +9/16\\
 & &
+  \frac{1}{2}\bigoplus_{j\in\mathbf{Z}^3}\!(\check{a}(j)\cdot \check{a}(j)\ + \ \check{e}(j)\cdot \check{e}(j)).
\end{eqnarray*}
 Meanwhile the transversality  equation (\ref{eq:transversality})
is the direct sum of the transversality  equations over  $\mathbf{C}\mathfrak{g}$.

The  operators $\widehat{H}_j:\ \mathcal{K}(\mathbf{C}\mathfrak{g}^*)\rightarrow
 \mathcal{K}(\mathbf{C}\mathfrak{g})$ over the finite-dimensional  space
 $\mathbf{C}\mathfrak{g}$ 
with the Weyl symbols 
\begin{equation}
\label{eq:Weylsymbols}
\sigma_\omega^{\widehat{H}_j}(a,e)\ =\ \frac{1}{2} \Big\| j\times \check{a}(j) - 
[ \check{a}(j)\stackrel{\times}{,}\check{a}(j)] \big)\Big\|^2\ +\ \frac{1}{2}\big(\check{a}(j)\cdot \check{a}(j)\big)\ + \ \check{e}(j)\cdot \check{e}(j)+\ 9/16.
\end{equation}
By (\ref{eq:Weylsymbols}), the operators $\widehat{H}_j$ and $\mathbf{N}$ are hypo-elliptic (see \textsc{\small shubin}\cite[Definition 25.1]{Shubin}). Hence, by \textsc{\small shubin}\cite[Theorem 26.3]{Shubin}, they have  countable orthonormal eigenbases in $\mathcal{S}(\mathbf{C}\mathfrak{g})$ with positive eigenvalues $\lambda_n(\widehat{H}_j)$  and 
$\lambda_n(\mathbf{N})$ converging to  $+\infty$ as $n\rightarrow +\infty$.

Let  $\mathcal{B}_j\subset\mathcal{K}(\mathcal{H}^*)$ be the eigenbases  of $\widehat{H}_j$ of the eigenvectors $v_{k,j}\in\mathcal{B}_j$.
Then  the finite  monomial products of $v_{j,k}$ are eigenvectors of the operator $\widehat{H}$ with the eigenvalue 
$\sum_{j,k}\lambda_{j,k}(\widehat{H}_j)$ (see  \textsc{small reed-simon}\cite[Theorem VIII.33]{Reed1}. 

Finally the finite monomial products together with a constant state form a countable  basis in $\mathcal{K}(\mathcal{H}^0)$. \qed

As  the proof shows, the spectrum  of $\widehat{H}$ is the set of  finite sums  of   $\widehat{H}_j$ eigenvalues.
(see \textsc{\small reed-simon}\cite[Chapter VIII]{Reed1}.

\end{document}